\documentclass[showpacs,preprintnumbers,amsmath,amssymb]{revtex4}

\usepackage{graphicx,color}
\usepackage{dcolumn}
\usepackage{bm}

\newcommand{\beq}{\begin{equation}}
\newcommand{\beqa}{\begin{eqnarray}}
\newcommand{\eeq}{\end{equation}}
\newcommand{\eeqa}{\end{eqnarray}}
\newcommand{\etal}{{\it et al.}}

\newcommand{\lsim}{\lesssim}
\newcommand{\gsim}{\gtrsim}

\newcommand{\lmk}{\left(}
\newcommand{\rmk}{\right)}

\newcommand{\lla}{\left\langle}
\newcommand{\p}{\partial}
\newcommand{\rra}{\right\rangle}
\newcommand{\so}{M_\odot}
\newcommand{\mch}{{\cal M}}
\newcommand{\rch}{{\cal R}}
\newcommand{\ach}{{\cal A}}
\newcommand{\fch}{{\cal F}}

\newcommand{\mrm}{\mathrm }
\newcommand{\cmdot}{C_{\dot{M}}}
\newcommand{\tzero}{{t0}}

\begin{document}

\preprint{KUNS-2319}
\preprint{YITP-11-11}

\title{ Probing the size of extra dimension with gravitational wave astronomy}

\author{Kent Yagi}

\affiliation{%
Department of Physics, Kyoto University,
   Kyoto, 606-8502, Japan
}%

\author{Norihiro Tanahashi}%
\affiliation{%
Department of Physics, University of California, Davis, CA 95616
}%

\author{Takahiro Tanaka}%
\affiliation{%
Yukawa Institute for Theoretical Physics,
  Kyoto University,
  Kyoto 606-8502, Japan
}%

\date{\today}

\begin{abstract}

In Randall-Sundrum II (RS-II) braneworld model, it has been conjectured
according to the AdS/CFT correspondence
that brane-localized black hole (BH) larger than the bulk AdS curvature
 scale $\ell$ cannot be static, and it is dual to a four dimensional BH emitting the Hawking radiation through some quantum fields.
In this scenario, the number of the quantum field species is so
 large that this radiation changes
the orbital evolution of a BH binary.
We derived the correction to the gravitational waveform phase due to this 
effect and estimated the upper bounds on $\ell$ by performing Fisher analyses. 
We found that DECIGO/BBO can put a stronger constraint than the current
 table-top result by detecting gravitational waves from small mass BH/BH
 and BH/neutron star (NS) binaries.
Furthermore, DECIGO/BBO is expected to detect 10$^5$ BH/NS binaries per year.
Taking this advantage, we found that DECIGO/BBO can actually measure $\ell$ down to $\ell=0.33 \mu$m for 5 year observation if we know that binaries are circular a priori.
This is about 40 times smaller than the upper bound obtained from the table-top experiment.
On the other hand, when we take eccentricities into binary parameters, the detection limit weakens to $\ell=1.5 \mu$m due to strong degeneracies between $\ell$ and eccentricities.
We also derived the upper bound on $\ell$ from the expected detection
 number of extreme mass ratio inspirals (EMRIs) with LISA and BH/NS
 binaries with DECIGO/BBO, extending the discussion made
 recently by McWilliams~\cite{mc}.
 We found that these less robust constraints are weaker than the ones from phase differences.
\end{abstract}

\pacs{Valid PACS appear here}
\maketitle

\section{INTRODUCTION}

Super-string theory suggests that our universe has more than four dimensions~\cite{polchinski} with extra dimensions being compactified in some way.
One very famous and simple way is the Kaluza-Klein compactification.
The size of extra dimension $\ell$ in this case is strongly
bounded from particle physics experiments as $\ell \le 10^{-16}$cm.
A new possibility opened after Arkani-Hamed \textit{et al}. 
proposed a braneworld model (the ADD
model)~\cite{arkani1,arkani2} (see also Ref.~\cite{antoniadis} for the string
realization of low scale gravity and braneworld models).  
They embedded a tension-less brane (on which we live) in a flat and
compact bulk spacetime and assumed ordinary matters 
to be localized on it.
In this case, gravitons are the only components that can propagate
through the bulk.
Since the law of gravity has been constrained only weakly by
experiments, the size of extra dimensions can be relatively large 
in this model. 
Moreover, the ADD model can give an alternative way to explain 
the hierarchy problem between the Planck scale
and the electroweak scale if the spacetime dimension
is 6 and the size of compactified bulk is 1mm.  

A different type of braneworld models
have been proposed by Randall and Sundrum~\cite{randall1,randall2}.
In their first model (RS-I model)~\cite{randall1}, a 
positive tension brane and a negative tension brane (on which we live) 
give boundaries of a five dimensional bulk space with a negative cosmological
constant.  
The unperturbed bulk is anti-de Sitter space whose metric 
is given by  
\beq
ds^2=e^{-2y/\ell} \eta_{\mu\nu}dx^{\mu}dx^{\nu}+dy^2,
\eeq
where $\eta_{\mu\nu}$ represents the Minkowski metric 
with $\mu,\nu$ indices running from 0 to 3, $x^{\mu}$ are the coordinates on the brane 
and $y$ is the coordinate of the extra dimension. 
$\ell$ is the AdS curvature length which characterizes the size of extra dimension.
The location of each unperturbed brane is specified 
by a $y=$constant surface, and $Z_2$ symmetry across the 
brane is assumed. 
This model can also solve the hierarchy problem by tuning the 
separation between two branes to be $\sim 37 \ell$.

In their second model (RS-II model)~\cite{randall2}, it is assumed that
we live on the positive 
tension brane. 
In this case, the position of the negative tension brane 
becomes almost irrelevant. 
Even if we send it to $y=\infty$, 
the extra dimension is still effectively compactified 
thanks to the exponential warp factor $e^{-2y/\ell}$ in the metric.
This model does not give any clue to 
the hierarchy problem, but it has a fascinating property: 
Although the model has a non-compact extra dimension, 
four dimensional general relativity is approximately reproduced 
on the positive tension brane. 
The gravitational potential between two masses $m$ and $M$ 
with separation $r$ on the brane becomes~\cite{garriga}
\beq
V=-\frac{GmM}{r}\left( 1+\frac{2}{3}\frac{\ell^2}{r^2} 
 +\cdots\right), \label{pot}
\eeq
with $G$ representing the effective gravitational constant. 
For the power-law corrected potential, the current table-top experiment
puts a constraint $\ell \le 14 \mu$m~\cite{adelberger}, 
which is derived from the
results obtained in Ref.~\cite{kapner}.
In this paper, we consider this RS-II model.

Next, we explain a non-trivial constraint on $\ell$ in the RS-II model 
that can be derived from the astrophysical observation of BHs.
So far, theoretically no brane-localized BHs larger than $\ell$ 
have been constructed either analytically or numerically.
By employing the AdS/CFT correspondence~\cite{maldacena,aharony} to
the brane-localized BHs, 
it has been conjectured that such BHs cannot be
static~\cite{emparan-conj,tanaka-conj} (see also
Ref.~\cite{hawking-brane} as an application of the AdS/CFT
correspondence to the RS-II braneworld scenario).
Applying the AdS/CFT correspondence, 
a five dimensional BH is considered to be dual to a four
dimensional BH associated 
with CFT fields. 
The latter system should evolve via Hawking emission from the BH.
According to the dictionary of the AdS/CFT correspondence, the number of
degrees of freedom of CFT is as large as
\beq
g_* = 15\pi \frac{\ell^2}{G}  \sim 10^{61}\left( \frac{\ell}{10\mu\mathrm{m}} \right)^2.
\eeq
%
This enormously large factor enhances the BH evaporation rate
considerably.
For a BH with mass $M$, the evaporation rate is evaluated as~\cite{emparan}
\beq
\frac{dM}{dt}
= -2.8\times10^{-7}
\left( \frac{1 M_{\odot}}{M} \right)^2 \left( \frac{\ell}{10\mu
\mathrm{m}} \right)^2 M_{\odot} \mathrm{yr}^{-1}
=:-\cmdot\left( \frac{\ell}{M} \right)^2 , \label{massloss}
\eeq
where $\cmdot$ was defined as the coefficient of the mass loss rate 
for later use.
This leads to the estimate of the lifetime 
of the BH, 
\beq
\tau\simeq 1.2\times 10^6 \left( \frac{M}{1 M_{\odot}} \right)^3 \left( \frac{10\mu \mathrm{m}}{\ell} \right)^2 \mathrm{yr}. \label{lifetime}
\eeq
Emparan \textit{et al}.~\cite{emparan} pointed out that primordial BHs, if detected, may put a strong bound on $\ell$.
Several constraints have already been obtained from astrophysical BHs
 by (i) estimating their masses and ages and by (ii) measuring the
orbital decay rates of BH binaries. 
For the former cases, Psaltis~\cite{psaltis} estimated the lower limit
on the age of the BH in the X-ray binary XTE J1118+480 to be 11Myr, 
which leads to $\ell \le 80\mu$m.
Gnedin \textit{et al}.~\cite{gnedin} 
also estimated the age of the BH in the extra-galactic globular 
cluster RZ2109, and obtained a 
conservative bound $\ell \le 10\mu$m.
For the latter cases, Johannsen \textit{et
al}.~\cite{johannsen1,johannsen2} focused on the upper bound on the
orbital decay rates of the X-ray binaries A0620-00 and XTE J1118+480, 
and placed bounds $\ell \le 161\mu$m and 
$\ell \le 970\mu$m, respectively.
%
The inclination $i$ of the binary A0620-00 was estimated 
as $i=51.0^{\circ}\pm 0.9$ by Cantrell \etal~\cite{cantrell}, and it leads to
the BH mass $M=6.6\pm 0.25 \so$. 
Since this is almost 2 times smaller than the one assumed in
Refs.~\cite{johannsen1, johannsen2}, this new results may put 
a slightly stronger constraint. 
Recently, 
Simonetti \etal~found that if the orbital decay rate of a
BH-pulsar binary is detected in future with the same accuracy as 30
years observation of PSR B1913+16~\cite{weisberg2004}, the 5 $\sigma$
upper bound on $\ell$ becomes 0.17$\mu$m~\cite{simonetti}. 

It is also possible to constrain $\ell$ from future gravitational wave
observations.
Inoue and Tanaka~\cite{inoue} derived 
the leading correction to the gravitational wave (GW) phase 
due to the modification in the gravitational potential 
mentioned in Eq.~(\ref{pot}).
By detecting GW signals from sub-lunar mass BH binaries with the 3rd
generation GW interferometers, they obtained a rather weak upper bound
on $\ell$. 

Recently, McWilliams~\cite{mc} estimated the possible constraints on $\ell$ in future by
using the Laser Interferometer Space Antenna (LISA)~\cite{danzmann} from 2
different observables; (i) the event rate of extreme mass ratio inspiral (EMRI)
and (ii) GW signal from a galactic BH/neutron star (NS) binary.
From the former estimate, 
it was claimed that $\ell$ can be constrained as $\ell \le
6\mu$m for a $(5+10^6)M_{\odot}$ binary 
if the predicted EMRI event rate in General
Relativity (GR)~\cite{gair} is correct. 
From the latter estimate, a monochromatic binary signal of a galactic
$(2+5)M_{\odot}$ BH/NS binary at $f=10^{-4}$Hz puts a constraint 
$\ell \le 22\mu$m
(see Ref.~\cite{wiseman} for the relativistic stars in RS-II model),
assuming that it is in the inspiral phase.
However, in order to probe the mass loss effect more robustly,
we need to determine the change in the orbital separation, which cannot
be made from a monochromatic signal.
Namely, we need to detect chirp (or anti-chirp) of GW signals from 
binaries.
In detecting these signals from a stellar mass BH/NS binary, the
Deci-Hertz Interferometer Gravitational Wave Observatory
(DECIGO)~\cite{decigo,kawamura2006} and the Big Bang Observatory
(BBO)~\cite{phinneybbo}, both having optimal sensitivities at 0.1-1Hz,
perform better than LISA.  

In this paper, we first derive 
the correction to the GW phase due to 
the
mass loss effect, whose frequency dependence behaves like ``-4PN'' correction (see Ref.~\cite{yunes} for a related work).
Then, we perform the matched filtering analyses and estimate the
possible constraints on $\ell$ by detecting GWs from BH binaries with LISA and DECIGO/BBO.
Since high BH/NS event rate of $O(10^5 \ \mrm{yr}^{-1})$ has been
predicted for DECIGO/BBO in GR~\cite{cutlerharms}, we can 
obtain a stronger constraint by performing statistical analyses. 
We also estimate the upper bounds on $\ell$ from the expected 
detection number of EMRI events for LISA and of BH/NS ones for DECIGO/BBO, extending the previous work by McWilliams~\cite{mc}. 

This paper is organized as follows.
In Sec.~\ref{sec-mc}, we review 2 ways of obtaining the constraints on
$\ell$ with LISA estimated by McWilliams~\cite{mc}.
In Sec.~\ref{sec-fisher}, we re-examine the constraints by
performing matched filtering analyses. 
First, we derive ``-4PN'' correction term 
appearing in the gravitational waveform phase.
Next, we describe the basics of the Fisher analysis and explain the
future planned space-borne interferometers that we use for our analyses.
Then, we evaluate the constraint obtained from a single event.
After that, we perform statistical analyses and show that the
constraints are improved. 
In Sec.~\ref{event}, we extend the discussions
made in Ref.~\cite{mc} for obtaining constraints from the number of
detection events.
We consider not only EMRI detection numbers with LISA but also BH/NS ones with DECIGO/BBO.
Finally in Sec.~\ref{conclusions}, we summarize our work and comment on
several issues that we did not take into account in this paper.
We also mention possible future works at the end. 
We take the present Hubble parameter as $H_0=72$km s$^{-1}$ Mpc$^{-1}$ and the
cosmological density parameters as 
$\Omega_m=0.3$ and $\Omega_{\Lambda}=0.7$.
Hereafter, we take the unit $G=c=1$ throughout this paper.

\section{Constraints with LISA obtained by McWilliams }
\label{sec-mc}

In this section, we review two ways to constrain the size of extra
dimension $\ell$ using LISA developed by McWilliams~\cite{mc}.
LISA will detect an almost monochromatic GW signal from a
galactic binary composed of a BH and a NS. 
The event rate for such a binary in the LISA frequency range 
is expected to be 1 yr$^{-1}$~\cite{nelemans}.
GW emission makes the orbital separation $a$ smaller (inspiral) with the orbital decay rate given by~\cite{flanagan}
\beq
\dot{a}_\mrm{GW} = -\frac{64}{5}\frac{\mu M_t^2}{a^3},
\eeq
where $M_t=m+M$ is the total mass of the binary and $\mu=mM/M_t$ is the reduced mass.
On the other hand, the BH mass loss effect makes $a$ larger (outspiral)
at the rate of
\beq
\dot{a}_H = -\frac{\dot{M}}{M_t} a, \label{rhawk1}
\eeq
%
where the mass loss rate is given by Eq.~(\ref{massloss}).
This is derived from the conservation of the specific orbital angular
momentum $j=\sqrt{M_t a}$ assuming that the radiation is emitted
isotropically in the rest frame of the BH~\cite{footnote_specific}.
There exists a critical separation $a_\mrm{crit}$ where $\dot{a}_\mrm{GW}$ and $\dot{a}_H$ balance.
If the separation is larger than $a_\mrm{crit}$, the mass loss effect
dominates over the GW emission and the separation gets larger while if
$a$ is smaller than $a_\mrm{crit}$, GW emission wins and the separation
gets smaller.
Typically, a galactic BH binary forms with its orbital period
$O$(days)~\cite{belczynski} whose GW frequency being slightly lower than
the LISA sensitivity band.
Therefore if its signal is detected at $f=10^{-4}$Hz, it means that GW
emission effect is dominating over the mass loss effect at this frequency.
The inequality $a(f=10^{-4}\mathrm{Hz}) \le a_\mrm{crit}$ leads to the
constraint
\beq
\ell \le 22 \left(\frac{M}{5 M_{\odot}} \right)^{3/2} \left(\frac{M_t}{7 M_{\odot}} \right)^{1/3} \left(\frac{m}{2 M_{\odot}} \right)^{1/2} \left(\frac{f}{10^{-4}\mathrm{Hz}} \right)^{4/3} \mu \mathrm{m},
\eeq
where the typical BH and NS masses are assumed 
to be $ M=5 M_{\odot}$ and $ m=2 M_{\odot}$, respectively. 


McWilliams also obtained a constraint on $\ell$ from the average EMRI
event rate $\lla \rch\rra_{\mathrm{EMRI}}$. 
The rate is estimated as~\cite{gair} 
\beq
\lla \rch \rra _{\mathrm{EMRI}}  \simeq \lmk \frac{M}{10^6\so} \rmk ^{3/8} \lmk \frac{5\so}{m} \rmk ^{1/2} \mrm{Gpc}^{-3} \mrm{yr}^{-1}. \label{emri}
\eeq
However, if we take into account the conjectured BH mass loss effect 
in the RS-II braneworld model, the evaporation time (rather than the age of the universe) also affects the event rate.
The modified estimate for the 
average EMRI event rate $\lla  \rch\rra_{\mathrm{H}}$ becomes
\beqa
\lla \rch \rra_{\mathrm{H}} & = & \lmk \frac{\tau}{10^{10}\mrm{yr}} \rmk \lla \rch \rra_{\mathrm{EMRI}}  \notag \\
        & = & 7.7\times 10^{-3} \lmk \frac{14\mu m}{\ell} \rmk^2 \lmk \frac{M}{10^6\so} \rmk^{3/8} \lmk \frac{m}{5\so} \rmk^{5/2} \mrm{Gpc}^{-3} \mrm{yr}^{-1}, \label{rh}
\eeqa
where the BH lifetime $\tau$ is given in Eq.~(\ref{lifetime}).
In Ref.~\cite{mc}, the author assumed that this event rate obeys the Poisson
probability distribution and estimated a constraint on $\ell$ that will be obtained if $\lla
\rch\rra_{\mathrm{EMRI}}$, the value predicted in GR, is
actually observed. 
However, here we point out that it is not the EMRI event rate 
but the detection number of EMRIs 
that obeys the Poisson probability distribution.  
We also have to take into account the large uncertainties in the event rate estimations. 
We study these issues later in Sec.~\ref{event}.  


\section{Constraints from the matched filtering analysis}
\label{sec-fisher}

In Sec.~\ref{sec-mc}, we mentioned that the typical galactic BH binary
forms with a GW frequency slightly lower than the LISA sensitivity band.
However, some of them may form with a frequency higher than
$f=10^{-4}$Hz~\cite{belczynski}.
Therefore, even if we detect a monochromatic binary signal at
$f=10^{-4}$Hz, we cannot immediately conclude 
that the binary separation is getting smaller.
What we need to detect is the changing rate of the separation which
leads to the variation of the GW frequency. 
In other words, we need to detect a chirping or anti-chirping signal.
In this section, we first derive the correction term in the GW phase due
to the mass loss effect.
Then, we evaluate the possible
constraint on $\ell$ from the matched filtering 
analysis using LISA or DECIGO/BBO.  
Throughout this paper, we assume that the binaries are quasi-circular.
We also neglect the spins of BHs and NSs.

\subsection{Binary Waveforms}

First we study the GW waveform from a binary composed of two BHs with
masses $M$ and $m$ (with $M\geq m$).
From Eqs.~(\ref{massloss}) and~(\ref{rhawk1}), the orbital separation
change due to the mass loss effect becomes
\beq
\dot{a}_H 
= \cmdot \frac{(m^2+M^2)\ell^2}{\mu^2 M_t^3} a. \label{rhawk}
\eeq
Then, from the total separation shift
$\dot{a}=\dot{a}_\mrm{GW}+\dot{a}_H$, the GW 
frequency shift becomes
\beqa
\dot{f} = \frac{\dot{\Omega}}{\pi} 
= \frac{96}{5}\pi^{8/3} \mathcal{M}^{5/3} f^{11/3} \biggl[ 1& - &\frac{5}{48}\cmdot\frac{1-2\eta }{\eta^3} \frac{\ell^2}{M_t^2} x^{-4}\notag \\
& - & \left( \frac{743}{336}+\frac{11}{4}\eta \right) x 
           +4\pi x^{3/2}  
           +\left( \frac{34103}{18144}+\frac{13661}{2016}\eta +\frac{59}{18}\eta^2 \right) x^2 \biggr], \label{fdot}
\eeqa
up to 2PN orders.
Here $\Omega\equiv M_t^{1/2}/a^{3/2}$ is the orbital 
angular velocity of the
binary and $x\equiv v^2\equiv (\pi M_t f)^{2/3}$ is the squared
velocity of the relative motion. 
We have also introduced 2 mass parameters, the symmetric mass ratio $\eta \equiv \mu/M_t$ and the chirp mass $\mch \equiv M_t \eta^{3/5}$.
The first term in the square bracket 
in Eq.~(\ref{fdot}) represents the leading Newtonian term.
The second one is the ``-4PN'' correction term due to the mass loss effect which we take up to $O(\ell^2)$.
The rest of the terms represent the higher PN contributions up to 2PN~\cite{berti}. 

Next, we integrate Eq.~(\ref{massloss}) to yield 
\beq
M\sim M_0\left( 1-\frac{3\cmdot\ell^2}{M_{0}^3}(t-t_0) \right)^{1/3} \sim M_{0}-\frac{\cmdot\ell^2}{M_{0}^2}(t-t_0),
\eeq
where the subscript 0 denotes the quantity at the time of coalescence.
The expression for $m$ is obtained just by replacing $M_0$ with 
$m_0$. 
Using this mass formula, we integrate Eq.~(\ref{fdot}) to obtain the time before coalescence $t(f)$ and the GW phase $\phi(f)=\int^{t(f)}_0 2\pi f dt$ as
\beqa
t(f)= t_0- \frac{5}{256}\mathcal{M}_0(\pi \mathcal{M}_0 f)^{-8/3}  \biggl[ 1&-& \frac{5}{1536}\cmdot C L x_0^{-4} 
      +\frac{4}{3}\left( \frac{743}{336}+\frac{11}{4}\eta_0 \right) x_0  \notag \\
      & -& \frac{32}{5}\pi x_0^{3/2} 
       +2\left( \frac{3058673}{1016064}+\frac{5429}{1008}\eta_0+\frac{617}{144}\eta_0^2 \right) x_0^2 \biggr], \label{tf}
\eeqa
and
\beqa
\phi (f)=\phi_0-\frac{1}{16} (\pi \mathcal{M}_0 f)^{-5/3} \biggl[ 1&-& \frac{25}{9984}\cmdot C L x_0^{-4} 
           +\frac{5}{3}\left( \frac{743}{336}+\frac{11}{4}\eta_0 \right) x_0  \notag \\
       &-&10\pi x_0^{3/2} 
       +5\left( \frac{3058673}{1016064}+\frac{5429}{1008}\eta_0+\frac{617}{144}\eta_0^2 \right) x_0^2 \biggr],
\eeqa
respectively.
Here, we introduced $L\equiv
\ell^2/M_{\tzero }^2$, 
$x_0\equiv (\pi M_{\tzero } f)^{2/3}$ 
and 
\beq
C  \equiv  \frac{3-26\eta_0 +34\eta_0^2}{\eta_0^4}. \label{c}
\eeq

The binaries of our interest satisfy the following conditions, $d \ln
A/dt \ll d \phi/dt$ and $d^2 \phi/dt^2 \ll (d \phi /dt)^2$, where $A$ is
the GW amplitude and $\phi$ is the phase in the time domain.
Then, using the stationary phase approximation~\cite{flanagan}, the gravitational waveform in the Fourier domain is given as 
\beq
\tilde{h}(f)=\frac{\sqrt{3}}{2}\mathcal{A} f^{-7/6}e^{i\Psi (f)}. \label{h-tilde}
\eeq
In this paper, we only keep 
the Newtonian quadrupole term for the amplitude and 
average it over the directions and the orientations of
the binaries~\cite{berti},
\beq
\ach=\frac{1}{\sqrt{30}\pi^{2/3}}\frac{\mathcal{M}_0^{5/6}}{D_L}, \label{amp}
\eeq
where $D_L$ is the luminosity distance.  
On the other hand, we keep the phase up to 2PN order. 
This is the so-called restricted 2PN waveform. 
The GW phase in the Fourier space is given as
\beqa
\Psi (f) &=& 2\pi ft(f)-\phi(f)-\pi /4 \notag \\
        & = & 2\pi ft_0-\phi_0-\pi /4+\frac{3}{128} (\pi \mathcal{M}_0 f)^{-5/3} \biggl[ 1-  \frac{25}{19968}\cmdot C L x_0^{-4}  
        + \left( \frac{3715}{756}+\frac{55}{9}\eta_0 \right) x_0 
                  -16\pi x_0^{3/2} \notag \\
          & & \qquad \qquad \qquad \qquad \quad + \left( \frac{15293365}{508032}+\frac{27145}{504}\eta_0+\frac{3085}{72}\eta_0^2 \right) x_0^2  \biggr]. \label{phase}
\eeqa
%
%
%
The second term in the bracket is the ``-4PN'' correction 
term in GW phase due to the mass loss effect.
Notice that $C$ in Eq.~(\ref{c}) gets larger as $\eta$ gets smaller.
This is because 
when $M_t$ is fixed, smaller $\eta$ means smaller
$m$, leading to (i) larger mass loss effect, and 
(ii) suppressed GW radiation. 
Note also that $C=0$ when $\eta=\frac{13-\sqrt{67}}{34}\doteqdot 0.14$.  

For a BH/NS binary, the orbital separation change 
becomes 
\beq
\dot{a}_H 
= \cmdot \frac{m^2 \ell^2}{\mu^2 M_t^3} a, 
\eeq
and Eq.~(\ref{fdot}) becomes 
\beq
\dot{f} = \frac{96}{5}\pi^{8/3} \mathcal{M}^{5/3} f^{11/3} \biggl[ 1 - \frac{5}{48}\cmdot\frac{(1-2\eta)-\sqrt{1-4\eta} }{2\eta^3} \frac{\ell^2}{M_t^2} x^{-4} + (\mrm{higher \ PN \ terms}) \biggr]. 
\eeq
Then, the coefficient $C$ changes to
\beq
C  =  \frac{(3-26 \eta_0 +34 \eta_0^2 )+(-3+20 \eta_0 ) \sqrt{1-4\eta_0}}{2 \eta_0^4}.
\eeq
%

\subsection{Fisher Analysis}

The detected signal $s(t)=h(t)+n(t)$ contains both the GW signal $h(t)$ and the noise $n(t)$.
We apply the matched filtering analysis to estimate how accurately we can determine the binary parameters
$\bm{\theta}$~\cite{flanagan,finn} with future planned space-borne GW
interferometers. 
We assume that the detector noise is stationary and Gaussian.
Then, the noise $n$ follows the probability distribution given by 
\beq
p(n) \propto \exp \left[-\frac{1}{2}(n|n)\right], \label{gauss}
\eeq
where the inner product is defined as
\beq
(A|B)=4 \mathrm{Re}\int ^{\infty}_{0}df \, \frac{\tilde{A}^{*}(f)\tilde{B}(f)}{S_n(f)}, \label{scalar-prod}
\eeq
and the quantities with tilde are the Fourier components.
$S_n(f)$ is the noise spectral density of \textit{each} interferometer.
The signal to noise ratio (SNR) $\rho$ using a detector with $N_\mrm{int}$ effective interferometers is defined by~\footnote{$N_\mrm{int}$ in Eqs.~(\ref{snr}), (\ref{prob}) and~(\ref{fisherinv}) were missing in the old and published versions. } 
\beq
\rho\equiv \sqrt{N_\mrm{int} (h|h)}.
\label{snr}
\eeq
%
%
%
Given a GW signal $s(t)$, the probability distribution
that the parameter $\bm{\theta}$ is the true
parameter set becomes 
\beq
p(h(\bm{\theta})|s)\propto \exp\left[ -\frac{1}{2} N_\mrm{int} \Gamma_{ij}\Delta\theta^i\Delta\theta^j \right],
\label{prob}
\eeq
where $\Gamma_{ij}$ is called the Fisher matrix defined by
\beq
\Gamma_{ij}\equiv \lmk \frac{\p h}{\p \theta^i} \Big| \frac{\p h}{\p \theta^j} \rmk. \label{fisher}
\eeq
Then, the determination error $\Delta \theta^i$ of the parameter $\theta^i$
is estimated as 
%
\beq
\Delta \theta^i\equiv \sqrt{\frac{(\Gamma^{-1})_{ii}}{N_\mrm{int}}}. \label{fisherinv}
\eeq
Note that $(\Gamma^{-1})_{ii}$ is the variance of the parameter $\theta^i$ using a single interferometer when all the other parameters have been marginalized.

\subsection{Detector Noise Spectrum}

In this subsection, we briefly explain 3 future planned space-borne GW
interferometers, LISA, DECIGO and BBO, and introduce their noise
spectra.
First, LISA is an all-sky monitor having a quadrupolar antenna
pattern~\cite{danzmann}.
It consists 3 drag-free spacecrafts in which free-falling mirrors are
contained.
These spacecrafts form an approximate 
equilateral triangle with the length of each side 5$\times 10^6$km.
They orbit the Sun 20$^{\circ}$ behind the Earth with the detector plane
tilting 60$^{\circ}$ with respect to the ecliptic. 
We can arbitrarily choose 2 out of 3 arms to form 1 interferometer.
Then, we can linearly combine 3 arms to form another interferometer
which corresponds to rotating the first interferometer by 45$^{\circ}$.
Therefore this triangular detector contains $N_\mrm{int}=2$ individual
interferometers~\cite{cutler1998}. 

\begin{figure}[t]
  \centerline{\includegraphics[scale=.45,clip]{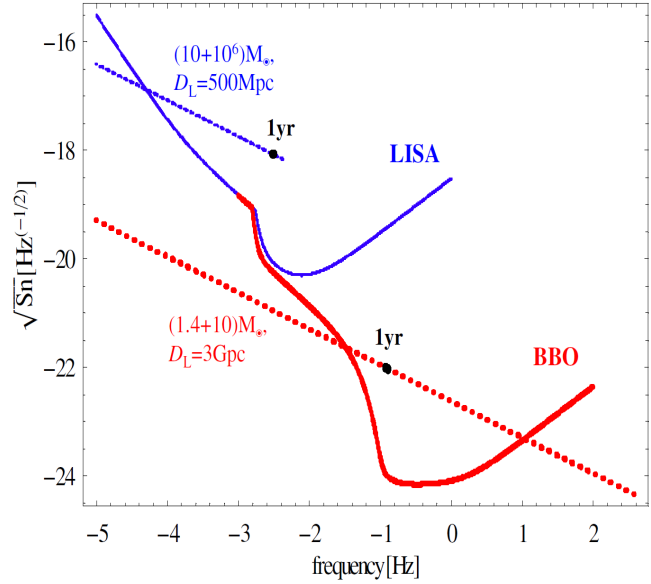} }
 \caption{\label{noise}
 The non sky-averaged noise spectral density for BBO (thick solid curve) and LISA (thin solid curve).
We also show the amplitudes of GWs from a $(10+10^6)M_{\odot}$ BH/BH binary at $D_L=500$Mpc (thin dotted line) and a $(1.4+10)M_{\odot}$ BH/NS binary at $D_L=3$Gpc (thick dotted line).  
Each dot labeled ``1 yr'' represents the frequency at 1 yr before the binary reaches ISCO. }
\end{figure}

For the noise spectrum of LISA, we follow the discussion by Barack
and Cutler~\cite{barack}.
The non sky-averaged instrumental noise spectral density for LISA is
given by~\footnote{Typos in the old and published versions have been corrected.}
\beq
S_{n,\mrm{LISA}}^{\mathrm{inst}}(f)=\left[ 9.2\times 10^{-52}\left( \frac{f}{1\mathrm{Hz}} \right)^{-4}+1.6\times 10^{-41}
                                                        +9.2\times 10^{-38} \left( \frac{f}{1\mathrm{Hz}} \right)^2 \right] \ \mathrm{Hz^{-1}}.
\eeq
In addition to that, there are confusion noises from the
galactic~\cite{nelemans} and the extra-galactic~\cite{farmer} white
dwarf (WD) binaries, 
\beqa
S_n^{\mathrm{gal}}(f) &=& 2.1\times10^{-45}\left(\frac{f}{1\ \mathrm{Hz}}\right)^{-7/3} \mathrm{Hz^{-1}}, \\
S_n^{\mathrm{ex-gal}}(f) &=& 4.2\times10^{-47}\left(\frac{f}{1\ \mathrm{Hz}}\right)^{-7/3} \mathrm{Hz^{-1}},
\eeqa
respectively.
Combining all, the total noise spectral density for LISA with a single interferometer becomes 
\beq
S_{n,\mrm{LISA}}(f)=\min\left[ \frac{S_{n,\mrm{LISA}}^{\mathrm{inst}}(f)}{\exp(-\kappa T_\mrm{obs}^{-1}dN/df)},\ 
            S_{n,\mrm{LISA}}^{\mathrm{inst}}(f)+S_n^{\mathrm{gal}}(f) \right] +S_n^{\mathrm{ex-gal}}(f). \label{noise-LISA}
\eeq
Here, $\kappa\simeq 4.5$ is the average number of frequency bins that
are lost when each galactic binary is fitted out and $T_\mrm{obs}$ is
the observation period.
$dN/df$ is the number density of galactic WD binaries per unit 
frequency given by~\cite{hughes}
\beq
\frac{dN}{df}=2\times10^{-3} \left(\frac{f}{1 \mathrm{Hz}} \right)^{-11/3} \mathrm{Hz}^{-1}.
\eeq
The lower and higher frequency ends of the LISA sensitivity band are
taken as $f_{\mrm{low}}=10^{-5}$Hz and $f_{\mrm{high}}=1$Hz,
respectively.
The noise spectrum of LISA is shown as a (blue) thin solid curve in
Fig.~\ref{noise}, together with an EMRI GW of $(10+10^6)\so$ at
$D_L=500$Mpc as a (blue) thin dotted line.

\begin{figure}[tbp]
  \includegraphics[scale=.5,clip]{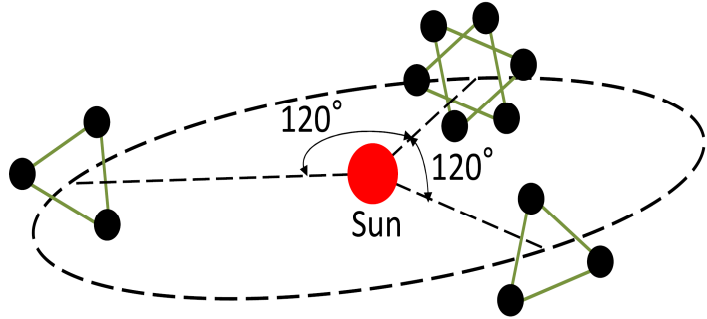} 
 \caption{\label{default} The configuration of DECIGO and BBO. There are 8 effective interferometers in total. }
\end{figure}

Next, we consider BBO~\cite{phinneybbo} and
DECIGO~\cite{decigo,kawamura2006}.
BBO consists of four triangular sets of detectors whose configuration is
shown in Fig.~\ref{default}.
This corresponds to $N_\mrm{int}=8$ individual interferometers.
Each triangle has an arm-length of 5$\times 10^4$km.
Its primary goal is to detect the primordial gravitational wave
background (PGWB) of $\Omega_\mrm{GW}=10^{-16}$.
Compared to LISA, BBO has an advantage in detecting the PGWB since the
WD/WD confusion noise will have a cutoff frequency at around 0.2
Hz~\cite{farmer} (see Refs.~\cite{cutlerharms,harms,yagiseto} for the
discussions of the NS/NS confusion noise).
In order to detect PGWB, it is necessary to perform correlation
analysis~\cite{flanagan-corr,allenromano}.
Therefore 2 of the 4 triangular detectors are located on the same site
forming a star of David.
The rest of the 2 detectors are placed far apart to enhance the angular
resolutions of the source location.
DECIGO has almost the same constellation as BBO.
The main difference is that while DECIGO is a Fabry-Perot type
interferometer, BBO is a transponder-type interferometer. 
DECIGO has arm-lengths of 1000km. 

The noise spectrum of BBO is given as follows.
The non sky-averaged instrumental noise spectral density for BBO is
obtained from Ref.~\cite{holz} as
\beq
S_{n,\mathrm{BBO}}^{\mathrm{inst}}(f)=
\left[
1.8\times 10^{-49} \left(\frac{f}{1 \mathrm{Hz}} \right)^2
                         +2.9\times 10^{-49}
                         +9.2\times 10^{-52}\left(\frac{f}{1 \mathrm{Hz}} \right)^{-4} \ \right]\mathrm{Hz^{-1}} .
\eeq
It has 20/3 times better sensitivity than the one for the sky-averaged
sensitivity~\cite{berti}.  
As for $S_n^\mrm{gal}$ and $S_n^\mrm{ex-gal}$, we multiply them by a
factor $\fch \equiv \exp\{-2\left({f}/{0.05\mathrm{Hz}} \right)^2\}$,
which corresponds to the high frequency cutoff for the white dwarf
confusion noises.
We also have to take into account the confusion noise from NS binaries, which is estimated as~\cite{cutlerharms, yagiseto}
\beq
S_n^{\mathrm{NS}}(f) = 1.3\times10^{-47}\left(\frac{f}{1\ \mathrm{Hz}}\right)^{-7/3}  \left(  \frac{\dot{n}_0}{10^{-6} \ \mathrm{Mpc}^{-3} \mathrm{yr}^{-1}} \right) \mathrm{Hz^{-1}},
\eeq
where $\dot{n}_0$ denotes current merger rate density of NS/NS binaries. 
Putting all together, the total noise spectral densities for BBO with a single interferometer becomes
\beq
S_{n,\mrm{BBO}}(f)=\min\left[ \frac{S_{n,\mrm{BBO}}^{\mathrm{inst}}(f)}{\exp(-\kappa T_\mrm{obs}^{-1}dN/df)},\ 
            S_{n,\mrm{BBO}}^{\mathrm{inst}}(f)+S_n^{\mathrm{gal}}(f) \fch (f) \right] +S_n^{\mathrm{ex-gal}}(f) \fch (f) + 0.001\times S_n^{\mathrm{NS}}(f) . \label{noise-BBO}
\eeq
The factor 0.001 in front of $S_n^{\mathrm{NS}}(f)$ represents our assumption of the fraction of GWs that cannot be removed after foreground subtraction~\cite{kent2}. 
(With this choice of the cleaning factor, the residual NS/NS foreground noise becomes below the instrumental noise.)
The lower and higher frequency ends of the BBO sensitivity band are set as $f_{\mrm{low}}=10^{-3}$Hz and $f_{\mrm{high}}=100$Hz, respectively.
The noise spectrum of BBO is shown as a (red) thick solid curve in Fig.~\ref{noise}.
We also show the amplitude of the GW signal from 
a BH/NS of $(1.4+10)\so$ at $D_L=3$Gpc as a (red) thick dotted line.
DECIGO has been proposed with 3-4 times less sensitive spectrum than
BBO.
However, this is not the fixed design sensitivity and there is a project
going on to improve the sensitivity to the same level as BBO.
Therefore, for the Fisher analyses below, we assume that both DECIGO and
BBO have the noise spectral densities shown in Eq.~(\ref{noise-BBO}).


\subsection{Numerical Setups}

In this subsection, we explain how we perform the numerical calculations
of the Fisher analyses.
We take 
\beq
\bm{\theta}=(\ln \mch_0, \ln \eta_0, t_0, \phi_0,D_L,L)
\eeq
as binary parameters, 
setting $t_0=\phi_0=0$ for the fiducial values.
We evaluate the determination errors of these
parameters, especially focusing on $L$, 
to estimate how strongly one can constrain or how accurately one can measure $\ell$. 
We assume that the observation starts $T$yr before coalescence.
While calculating the Fisher matrix, we perform the derivative of
$\tilde{h}(f)$ with respect to $\bm{\theta}$ analytically.
We take the integration range of $(f_{\mathrm{in}},f_{\mathrm{fin}})$
with 
\beq
f_{\mathrm{in}}=\max \bigl\{ f_{\mathrm{low}},  f_{T\mrm{yr}} \bigr\}, \qquad
f_{\mathrm{fin}}=\min \bigl\{ f_{\mathrm{high}},  f_{\mathrm{ISCO}} \bigr\}.
\eeq
Here $f_{T\mrm{yr}}$ is the frequency at the time $T$yr 
before the binary reaches the innermost stable circular orbit (ISCO), which is given as
\beq
f_{T\mrm{yr}}=5.5 \times 10^{-2} \left[ \left(\frac{\mathcal{M}_0}{10
M_{\odot}}\right)^{-5/8} \left( {T-t(f_\mrm{ISCO})\over 1 \mrm{yr}} \right)^{-3/8} \right] \mrm{Hz}, \label{fT}
\eeq
with $t(f)$ given in Eq.~(\ref{tf}).
Here, we only take into account the leading contribution from GR.
Higher PN contributions and the modification from mass loss effect on $f_{T\mrm{yr}}$ are not important for our analysis. 
\beq
f_{\mathrm{ISCO}}=\frac{1}{6^{3/2}\pi M_{\tzero }}  =4.3\times 10^{2} \lmk \frac{10 \so}{M_{t0}} \rmk \mrm{Hz}
\eeq
is the frequency at ISCO. 
We performed the numerical integration with the Gauss-Legendre 
routine GAULEG~\cite{numerical}.
This quadrature uses the zero points of the $n$-th Legendre polynomials
as the abscissas and the integrand can be calculated exactly up to
(2$n$-1)-th order.
We take $n=400$ for the analyses of single binary GWs in
Sec.~\ref{single} and $n=2000$ for the ones with statistical analyses of
10$^5 \ \mrm{yr}^{-1}$ event rate in Sec.~\ref{sec_stat}. 

We use the Gauss-Jordan elimination for inverting the Fisher
matrix~\cite{numerical}.
In order to make sure that the inversion being performed correctly, we
first normalized the diagonal components of the Fisher matrix to 1.
Then, we take the inversion and convert it to the inversion of the
original Fisher matrix (see Appendix C in Ref.~\cite{kent}).
We checked that our inversion has succeeded by simply multiplying the
inversed Fisher matrix with the original one and see how close the
result is to the identity matrix $\delta_{ij}$.

\subsection{Constraints from Single Binary GWs}
\label{single}

In this subsection, we show the results for the GW signal from a single binary, 
first when we use LISA and next DECIGO/BBO. 
Here, we take the fiducial value as $L=0$ so that we estimate the constraint on $\ell$ assuming that GR is the correct theory.
From Eq.~(\ref{fisherinv}), one can easily 
relate the constraint on $\ell$ 
to the Fisher matrix as 
\beq
\ell \le \ell_u \equiv (\Gamma^{-1})_{LL}^{1/4} M_{\tzero }, \label{upper}
\eeq
with $\ell_u$ denoting the upper bound on $\ell$.

\begin{figure}[t]
  \centerline{\includegraphics[scale=.7,clip]{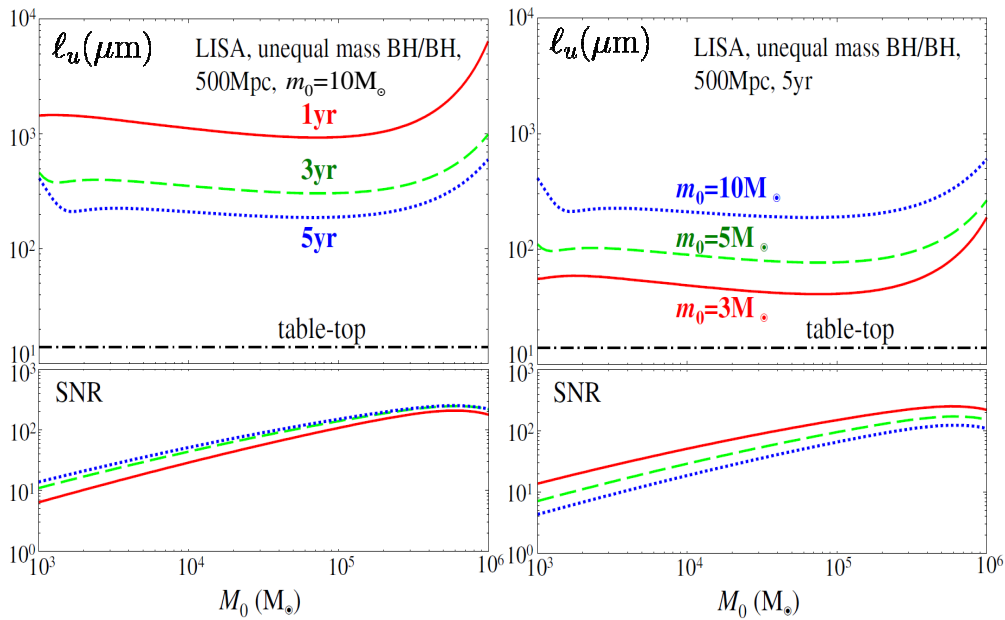} }
 \caption{\label{lisa}
(left) The upper bounds on $\ell$ (Eq.~(\ref{upper})) by detecting GWs from BH/BH binaries
 at $D_L=500$Mpc with LISA. 
We set $m_0=10\so$ and vary the values of $M_0$.
The (red) solid, (green) dashed and (blue) dotted curves represent the
bounds obtained from 1 yr, 3 yr and 5 yr observations, respectively.
The horizontal black dotted-dashed line shows the upper bound $\ell \le
 14\mu \mrm{m}$ obtained from the table-top
 experiment~\cite{adelberger}.
SNR for various mass binaries with each observation period is also shown
 at the bottom. 
(right) We show the 5 yr results for $m_0=3 \ \so$, $m_0=5 \ \so$ and $m_0=10 \ \so$ with the (red) solid, (green) dashed and (blue) dotted curve, respectively.  
Again, SNR for each binary is shown at the bottom.}
\end{figure}

\subsubsection{LISA}

Figure~\ref{lisa} shows the constraints on $\ell$ from unequal mass BH/BH binaries with LISA.
We place the binaries at $D_L=500$Mpc.
The detection rate of $(10+10^6)\so$ can be estimated from
Eq.~(\ref{emri}) as $\approx $1 yr$^{-1}$.
First, we set $m_0=10\so$ and vary $M_0$ from $10^3\so$ to $10^6\so$.
The results are shown in the left panels of Fig.~\ref{lisa} for
observation period of 1 yr, 3 yr and 5 yr, together with the table-top
constraint of $\ell \le 14\mu$m~\cite{adelberger}.
We also show the SNR of each binary at the bottom.
From the upper panel, we see that the upper bounds on $\ell$ from BH/BH
binaries with a large mass ratio have very weak dependence on $M_0$,
although SNR is increasing as $M_0$ gets larger.
This is simply because ``-4PN'' correction term in GW phase is smaller for larger $M_{\tzero }$. 
In fact, we can easily estimate that the 
upper bound on $\ell$ from EMRI scales as $M_0^{-1/8}$ as follows.

From Eqs.~(\ref{h-tilde})-(\ref{phase}), we get 
\beq
\frac{\p \tilde{h}}{\p L}  \propto  \ach f^{-7/6} C (\pi M_{\tzero } f)^{-8/3} (\pi \mch_{0} f)^{-5/3}  
                                \propto  M_{\tzero }^{-7/2} \eta_0^{-9/2} f^{-11/2}, 
\eeq
where we have used $C \propto \eta_0^{-4}$ for $\eta_0 \ll 1$.
For the BH/BH binaries considered here, $f_{T\mrm{yr}}$ ranges from $10^{-3}-10^{-2}$ Hz.
In this frequency range, the noise spectral density of LISA 
can be approximated by a constant. 
From these estimates with Eq.~(\ref{fisher}), 
the $L$-$L$ component of the Fisher matrix becomes
\beq
\Gamma_{LL} \propto \bigg| \frac{\p \tilde{h}}{\p L} \bigg|^2 \frac{f}{S_n(f)}\Bigg|_{f=f_{T\mrm{yr}}}
                  \propto M_{\tzero }^{-7} \eta_0^{-9} f_{T\mrm{yr}}^{-10}
                  \propto M_{\tzero }^{-3/4} \eta_0^{-21/4} T^{15/4},
\eeq
where we have used Eq.~(\ref{fT}).
If we can neglect the degeneracy between $L$ and the 
other binary parameters, 
the upper bound on $\ell$ follows from Eq.~(\ref{upper}) as
\beq
\ell_u \sim \ell_u ^{\mrm{(uncor)}} \equiv (\Gamma_{LL})^{-1/4} M_{\tzero } \propto M_{\tzero }^{19/16} \eta_0^{21/16}T^{-15/16}. \label{uncor}
\eeq
For the case of EMRI, $M_{\tzero }\sim M_0$ and $\eta_0\sim
\frac{m_0}{M_0}$. 
Then, we have
\beq
\ell_u ^\mrm{(uncor)} \propto  M_{0}^{-1/8} m_0^{21/16}T^{-15/16}. \label{lu}
\eeq
We find that $\ell_u$ depends only weakly on $M_0$.
Also, the scaling of $T^{-15/16} $ is roughly consistent 
with the upper left panel of Fig.~\ref{lisa}. 

Next, we fix the observation period as 5 yr and estimate $\ell_u$ for
$m_0=3\so$, $5\so$ and $10\so$. 
The results are shown in the right panels of Fig.~\ref{lisa}, together with SNRs at the bottom.
The $m_0$ dependence of $\ell_u$ is consistent with 
our simple estimate above, $\ell_u\propto m_0^{21/16}\sim m_0^{1.3}$.  
From Fig.~\ref{lisa}, we see that constraint on 
$\ell$ from BH/BH observation with LISA is 
weaker than the one from the current table-top experiment. 
Also we note that the binaries with relatively small $M_0$ may have too
small SNRs for applying Fisher analyses~\cite{vallisneri}.
In this case, thinking of binaries closer to us (though the event rate reduces to
less than 1 yr$^{-1}$), we can give a constraint scaling as 
$\ell_u \propto D_L^{1/2}$.

\begin{figure}[t]
  \centerline{\includegraphics[scale=.3,clip]{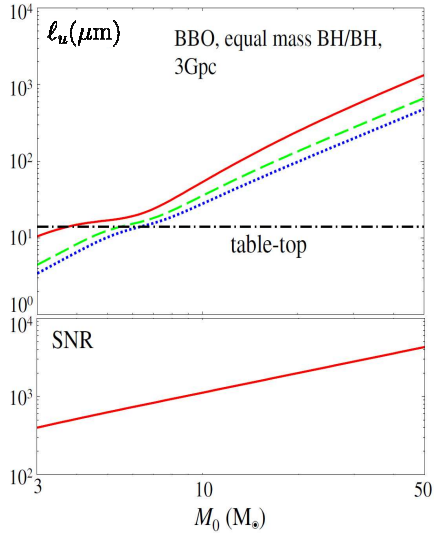} }
 \caption{\label{bbo-equal}
The upper bounds on $\ell$ by detecting GWs from equal mass BH/BH binaries at $D_L=3$Gpc with DECIGO/BBO. 
The meaning of each line is the same as the one in the left panels of Fig.~\ref{lisa}. 
Here only 1 curve is shown for the SNR since 1 yr, 3 yr and 5 yr results are almost indistinguishable.  }
\end{figure}

\subsubsection{DECIGO/BBO}

Here we examine the estimate of the constraint 
obtained by using DECIGO/BBO.
We fix the binary distance as $D_L=3$Gpc.
In Fig.~\ref{bbo-equal}, we show the upper bounds on $\ell$ from equal
mass BH/BH binaries with observation periods of 1 yr, 3 yr and 5 yr.
The meaning of each line is the same as in the left panels of
Fig.~\ref{lisa}.
Since we know that smaller mass binaries put stronger constraints
(because the evaporation times are smaller), we only vary $M_0$ from
3$\so$ to 50$\so$.
The results show that, by detecting GWs from a small mass binary, 
DECIGO/BBO can put a more stringent constraint on $\ell$
than the table-top experiment. 
For example, 1 yr (5 yr) observation of a (3+3)$\so$ BH/BH binary 
at 3 Gpc can put a constraint $\ell \le 10 \mu$m ($3.4 \mu$m).

\begin{figure}[t]
  \centerline{\includegraphics[scale=.7,clip]{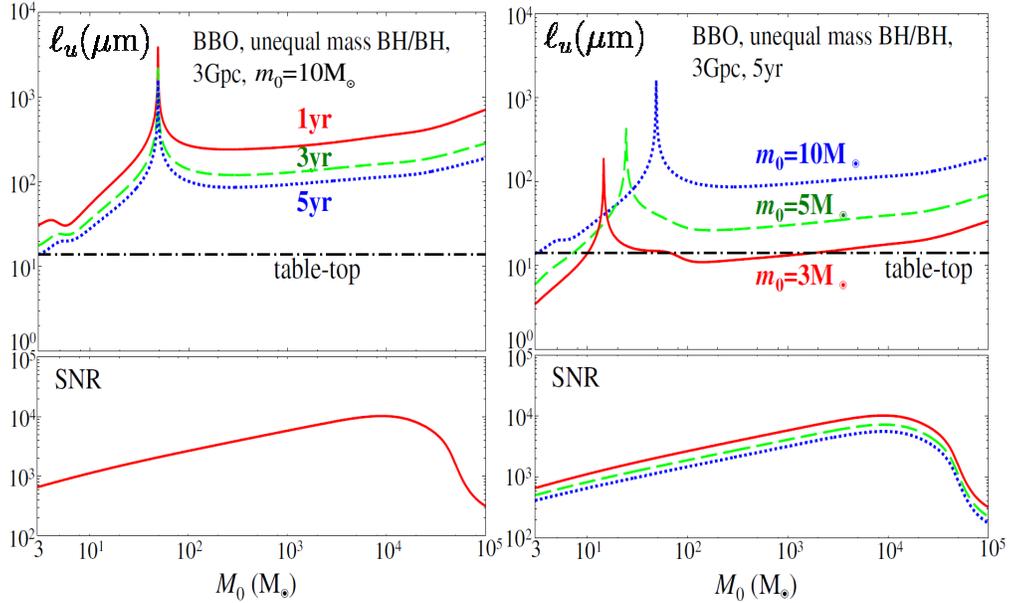} }
 \caption{\label{bbo-unequal}
The upper bounds on $\ell$ by detecting GWs from BH/BH binaries at $D_L=3$Gpc with DECIGO/BBO. 
The meaning for each line is the same as in Fig.~\ref{lisa}.
We set $m_0=10\so$ for the left panels and $T_\mrm{obs}=5$yr for the right panels. }
\end{figure}

Next, we consider unequal mass BH/BH binaries.
First, we fix $m_0=10\so$ and vary $M_0$ from $3\so$ to $10^5\so$ with
observation periods of 1 yr, 3 yr and 5 yr.
The values of $\ell_u$ obtained are displayed in the upper left panel of
Fig.~\ref{bbo-unequal}.
Here, the peaks at $M_0\sim 50\so$ correspond to $\eta_0 \sim 0.14$ where $C\sim 0$.
For binaries with large $M_0$, $\ell_u$ is almost independent of $M_0$,
which is the same feature explained in Fig.~\ref{lisa}.
In the right panels of Fig.~\ref{bbo-unequal}, we fix 
the observation period
as 5 yr and take $m_0=3\so$, $m_0=5\so$ and $m_0=10\so$. 
Again, only small mass binaries can put a stronger constraint than the
table-top experiment.

\begin{figure}[t]
  \centerline{\includegraphics[scale=.3,clip]{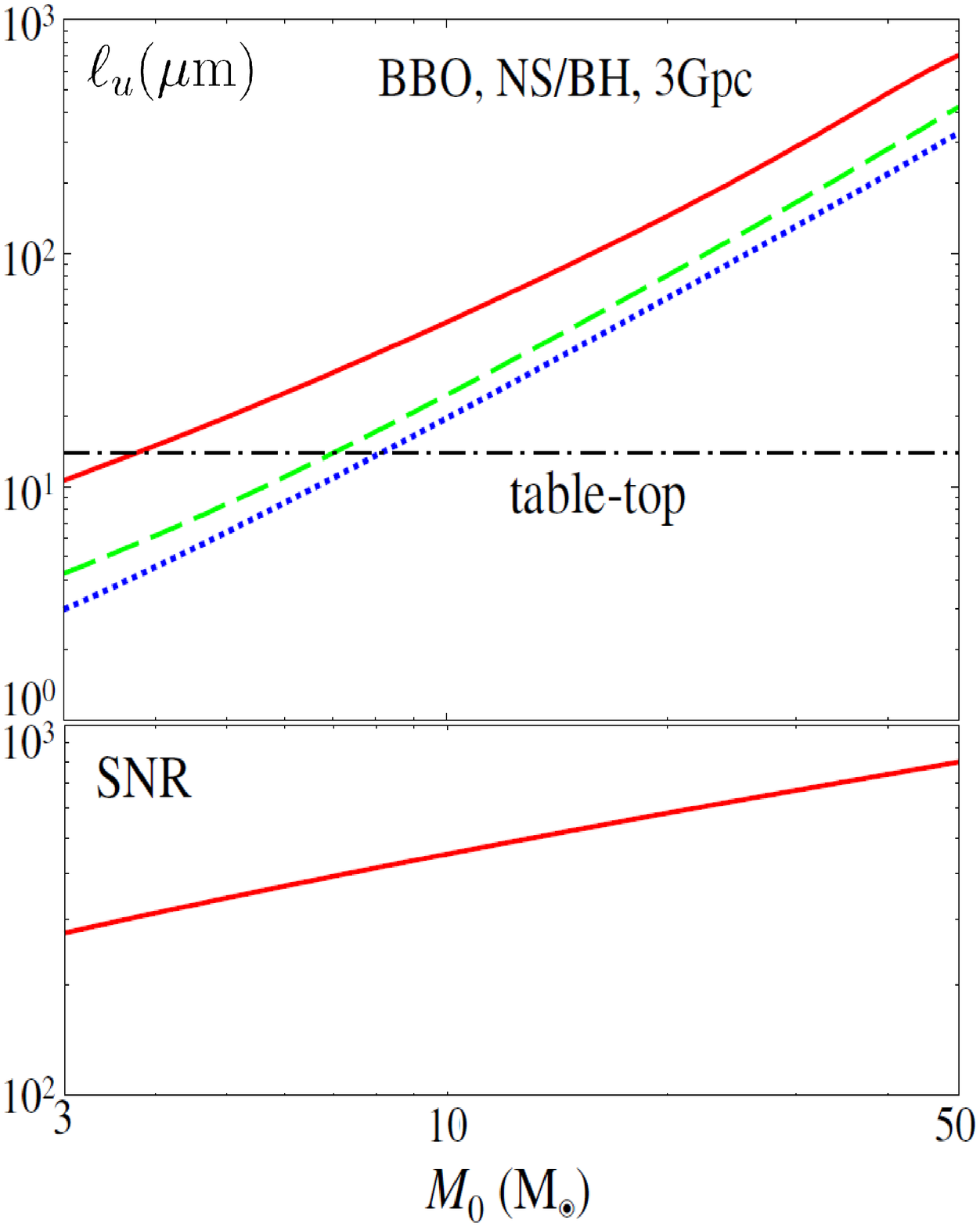} }
 \caption{\label{bbo-nsbh}
(Top) The upper bounds on $\ell$ by detecting GWs from BH/NS binaries at $D_L=3$Gpc with DECIGO/BBO. 
The meaning for each curve is the same as the one in the left panels of Fig.~\ref{lisa}.
We set the NS mass to $m_0=1.4\so$. }
\end{figure}

In Fig.~\ref{bbo-nsbh}, the upper bounds on $\ell$ from BH/NS binaries
with $m_0=1.4\so$ are shown. 
We take observation periods of 1 yr, 3 yr and 5 yr and vary $M_0$ from 3$\so$ to 50$\so$.
We see that some binaries put stronger constraint compared to the
table-top experiment and the features of this figure are similar to the
ones of Fig.~\ref{bbo-equal}.

\subsection{Statistical Analysis of 10$^5$ events/yr}
\label{sec_stat}

\begin{figure}[t]
  \centerline{\includegraphics[scale=.5,clip]{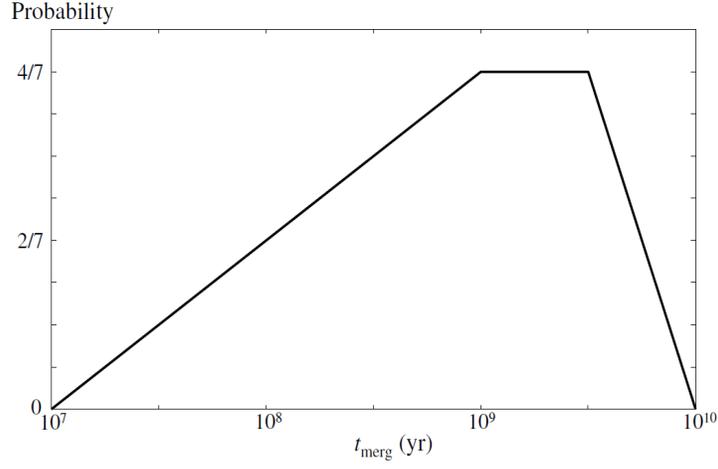} }
 \caption{\label{tmerg}
Probability distribution of NS/BH merger time $t_\mrm{merg}$~\cite{belczynski}. }
\end{figure}

It is expected that DECIGO/BBO will detect BH/NS GWs with the event rate
of $10^5$ yr$^{-1}$. 
Following Ref.~\cite{kent2}, we consider the 
enhancement of the parameter determination accuracy 
by performing statistical analyses that make use of a 
large number of events. 
First, we consider a BH/NS binary at a redshift $z$ 
with the BH mass (at the time of coalescence) being $M_0$. 
In order to take into account the complete degeneracy between mass and
redshift, we replace the mass parameters in the Fisher analyses 
to the redshifted ones 
$m_0\rightarrow m_{0z}\equiv (1+z)m_0$ and $M_0\rightarrow M_{0z}\equiv (1+z)M_0$.
The way how $L$ is defined in terms of the original total mass, 
$L\equiv \ell^2/M_{\tzero}^2$, is unaltered.
Then, the total variance $\sigma^2_{\ell^2}$ of the parameter $\ell^2$ is given by
double integration of $z$ and $M_0$ as
\beq
\sigma_{\ell^2}^{-2}=T_\mrm{obs} \int^{M_{0,\mrm{max}}}_{M_{0,\mrm{min}}}
\int^{\infty}_{0} 4\pi [a_0 r(z)
]^2\dot{n}_L(z,M_0)\frac{d\tau}{dz}[\sigma_{L}(z,M_0)M_{t0}^2]^{-2} dz
dM_0. \label{statistics}
\eeq
Here $a_0$ represents the current scale factor and $r(z)$ is the comoving distance to the source given as~\cite{cutlerharms}
\beq
a_0 r(z) =\frac{1}{H_0} \int ^z_0 \frac{dz'}{\sqrt{\Omega_m (1+z')^3+\Omega_{\Lambda}}}. \label{a0rz}
\eeq
$\tau$ is the proper look back time of the source and $\frac{d\tau}{dz}$ is given as~\cite{cutlerharms}
\beq
\frac{d\tau}{dz} = \frac{1}{H_0(1+z)\sqrt{\Omega_m (1+z)^3+\Omega_{\Lambda}}}. \label{dtaudz}
\eeq
$\dot{n}_L(z,M_0)=f(M_0) f_L(M_0) \rch R(z)$~\cite{cutlerharms} shows the BH/NS
merger rate at redshift $z$ where $\rch$ is the estimated merger rate at today in GR and 
\begin{eqnarray}
R(z)=\left\{ \begin{array}{ll}
1+2z & (z\leq 1) \\
\frac{3}{4}(5-z) & (1\leq z\leq 5) \\
0 & (z\geq 5) \\
\end{array} \right.
\end{eqnarray}
encodes the time evolution of this rate~\cite{footnote}.
$f(M_0)$ denotes the mass function of the BH/NS binaries in GR normalized 
to satisfy $\int^{M_{0,\mrm{max}}}_{M_{0,\mrm{min}}} f(M_0) dM_0=1$ where
$M_{0,\mrm{min}}$ and $M_{0,\mrm{max}}$ represent the minimum and
maximum values of $M_0$ in the distribution, respectively.
$f_L(M_0)$ is the reduction rate of the total number of merging BH/NS
binaries with the BH mass being $M_0$ in RS-II model with $L$ compared 
to the case of GR.
$f_L(M_0)$ is estimated by calculating the probability that BH/NS
binaries merge within the BH evaporation time $\tau$ (Eq.~(\ref{lifetime})). 
For the probability distribution of the binary merger time
$t_\mrm{merg}$, we use the one shown in Fig.~\ref{tmerg} which is a
simplified fitting of the result shown in Fig.~8 of Ref.~\cite{belczynski}.
We here assume that this probability distribution does not depend on the BH masses, 
from which we obtain rather conservative results (see Sec.~\ref{conclusions} for further discussions). 

\begin{figure}[t]
  \centerline{\includegraphics[scale=.5,clip]{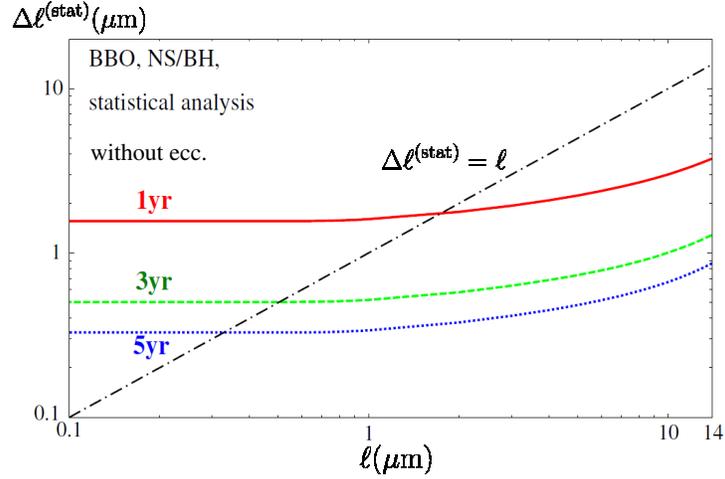} }
 \caption{\label{stat}
The detection errors of $\ell$ (Eq.~(\ref{det_error})) with statistical analysis for  1 yr (red
 solid), 3 yr (green dashed) and 5 yr (blue dotted) observations of BH/NS
 binaries with DECIGO/BBO.
The (black) dotted-dashed line represents $\Delta \ell^{(\mrm{stat})}=\ell$.
If $\Delta \ell^{(\mrm{stat})}$ comes below this line, $\ell$ can be measured.}
\end{figure}

We set $m_0=1.4\so$ and following Fig.~2 of Ref.~\cite{belczynski}, we
take the flat mass distribution between $M_{0,\mrm{min}}=3\so$ and
$M_{0,\mrm{max}}=13\so$.
We choose $\rch=10^{-7}$ Mpc$^{-3}$yr$^{-1}$ for our fiducial value (see
Sec.~\ref{event} for further details).
The determination error of $\ell$ for the statistical analysis, defined as 
\beq
\Delta \ell^{(\mathrm{stat})} \equiv \sqrt{\sigma_{\ell^2}}, \label{det_error}
\eeq
is shown against $\ell$ in Fig.~\ref{stat} for various observation periods of 1 yr (red solid), 3 yr (green dashed) and 5 yr (blue dotted).
As we lower $\ell$, $\Delta \ell^{(\mrm{stat})}$ becomes flat. 
This is because the BH evaporation time $\tau$ becomes larger than the
age of the universe for small $\ell$, meaning that the number of BH/NS
binaries detected is the same as the one in GR, irrespective of the
value of $\ell$. 
If $\Delta \ell^{(\mrm{stat})}$ is  below the dotted-dashed line that 
corresponds to $\Delta \ell^{(\mrm{stat})}=\ell$, $\ell$ can be detected.
From this figure, we see that DECIGO/BBO can measure $\ell$ down to $\ell
=1.8\mu$m (1 yr), $\ell=0.50\mu$m (3 yr) and $\ell=0.33\mu$m (5 yr). 
This is about 40 times stronger than the current table-top result~\cite{adelberger}.
On the other hand, if the curve of $\Delta \ell^{(\mrm{stat})}$ is 
above the line of $\Delta \ell^{(\mrm{stat})}=\ell$, 
%
we can not detect such small $\ell$ but only give constraints on it.
The constraints on $\ell$ obtained in this manner are summarized in Table~\ref{table-stat} when GR is correct (second row) and when RS-II is correct (fourth row).
If GR is the correct theory, the upper bound on $\ell$ becomes the ones shown in the second row of Table~\ref{table-stat}.
Note that the upper bound with 1 yr observation ($\ell=1.6 \mu$m) is slightly stronger than the detection limit of $\ell =1.8\mu$m.
This difference exists because if RS-II is the correct theory, some of BHs
may have evaporation time shorter than the age of the universe which
reduces the detection rate of binaries, leading to the weaker limit. 
The reason for this difference being small is because 
$\Delta \ell^{(\mrm{stat})}$ is only weakly dependent on the number of binaries as $\Delta \ell^{(\mrm{stat})}
\propto (\mrm{number \ of \ binary})^{1/4}$. 
For 3 yr or 5 yr observation, the upper bound is the same as the detection limit
value.
This is due to the fact that, if $\ell=0.50\mu$m, the evaporation time of
3$\so(=M_{0,\mrm{min}})$ BH is about 10$^{10}$ yr.
This means that the detection rates are 
the same in both RS-II and in GR, which leads to the same upper bound.  

\begin{table}[t]
\caption{\label{table-stat} 
DECIGO/BBO constraints on $\ell$ from a single $(1.4+3)\so$ BH/NS binary at $D_L=$3Gpc (first row) and from statistical analyses assuming that GR is the correct theory with (second row) and without (third row) taking eccentricity as a parameter.
We also show the detection lower limit on $\ell$ with DECIGO/BBO assuming that RS-II model is the correct theory with (fourth row) and without (fifth row) taking eccentricity as a parameter.
For the statistical analyses, we assume a flat BH mass distribution between $3\so$ and $13\so$ for
 BH/NS binaries with the total detection event rate of 10$^5$ yr$^{-1}$ for 
 DECIGO/BBO.
If RS-II is the correct theory, we cannot measure $\ell$ for 1 yr observation with taking eccentricity as a parameter since the determination error $\Delta\ell^{\mrm{(stat)}}$ is always larger than the fiducial value of $\ell$.}
\begin{center}
\begin{tabular}{|c||c|c|c|}  \hline
 obs. period (yr) & 1 yr & 3 yr & 5 yr  \\ \hline 
 single, GR, without ecc. ($\mu$m) &  11  & 4.3  & 3.0   \\ \hline
 statistical, GR, without ecc. ($\mu$m) & 1.6 & 0.50  &  0.33  \\ 
 statistical, GR, with ecc. ($\mu$m) & 6.5 & 2.2  &  1.4  \\ \hline
 statistical, RS-II, without ecc. ($\mu$m) & 1.8 & 0.50  & 0.33   \\ 
 statistical, RS-II, with ecc. ($\mu$m) & - & 2.7  & 1.5   \\
\hline  
\end{tabular}
\end{center}
\end{table}

\subsection{Effects of Eccentricities}

\begin{figure}[t]
  \centerline{\includegraphics[scale=.5,clip]{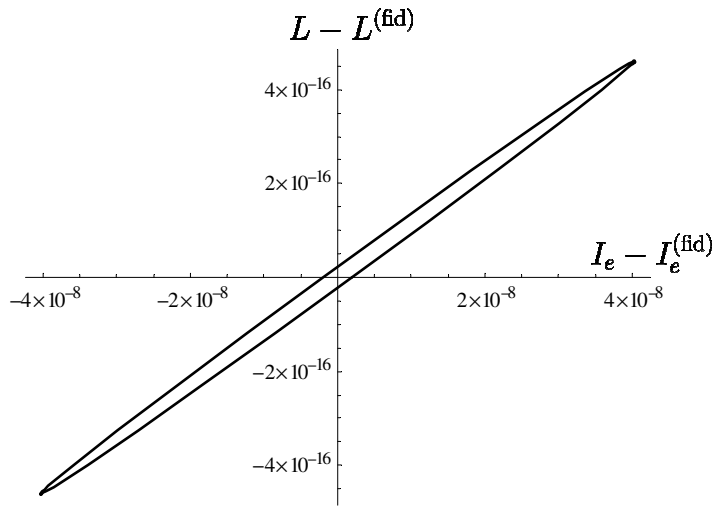} }
 \caption{\label{ecc}
The error contour showing 95$\%$ confidence level on the $I_e$-$L$ plane for a $(1.4+10)M_{\odot}$ BH/NS binary at $D_L=3$Gpc, such that the fiducial values lie at the center of the ellipse.
It is assumed that the observation period is 1 yr with DECIGO/BBO and we take $I_e^{\mrm{(fid)}}=L^{\mrm{(fid)}}=0$ for our fiducial values. }
\end{figure}

Up to now, we assumed that the binaries are quasi-circular.
This is because eccentricity $e$ decreases as $e\propto
f^{-19/18}$~\cite{peters} 
so that it is expected to be negligible by the time when binary GWs come 
into the observation band. 
However, some binaries, especially EMRIs and intermediate mass ratio
inspirals (IMRIs), may have non-negligible amount of $e$ even at 1 yr
before they reach ISCO.
For example, based on the results obtained by Hopman and
Alexander~\cite{hopman}, two of the authors of this paper 
estimated the eccentricity of a (1.4+10$^3$)$\so$ BH/NS binary 
typically as large as $e=0.026$ at $f=f_{1\mrm{yr}}$~\cite{kent}. 
Therefore we need to include eccentricity into binary parameters for more realistic analysis.
When averaged over 1 orbital period, the mass loss effect does not change $e$~\cite{simonetti}.
Therefore the leading contribution of eccentricity to the binary
waveform $\delta\Psi _e$ is unaltered~\cite{kent}:
\beq
\delta\Psi _e=-\frac{3}{128}(\pi \mathcal{M}_0f)^{-5/3}\frac{2355}{1462}I_e x_0^{-19/6},
\eeq
where $I_e\equiv (\pi M_{\tzero })^{19/9}e_c^2f_c^{19/9}$ is the
dimensionless asymptotic eccentricity invariant with $f_c$ and $e_c$
being an arbitrary reference frequency and the eccentricity at 
that frequency, respectively. 
In Fig.~\ref{ecc}, we show the 
95$\%$ confidence level contour in $I_e$-$L$
plane for a (1.4+10)$\so$ NS/BH binary
with the fiducial parameters, $I_e=0$, $\ell=0$ and an observation
period of 1 yr with DECIGO/BBO.
We see that there is strong degeneracy between these 2 parameters.
Unfortunately, we cannot obtain stronger constraint on $\ell$ even if we consider only the right half of Fig.~\ref{ecc}.
This means that prior information of $I_e>0$ does not affect the bound on $\ell$.
In the third row of Table~\ref{table-stat}, we show the constraints on $\ell$ obtained from statistical analyses assuming that GR is the correct theory and taking eccentricity into account.
These are about 4 times weaker than the ones without taking eccentricities into account (second row).

\begin{figure}[t]
  \centerline{\includegraphics[scale=.5,clip]{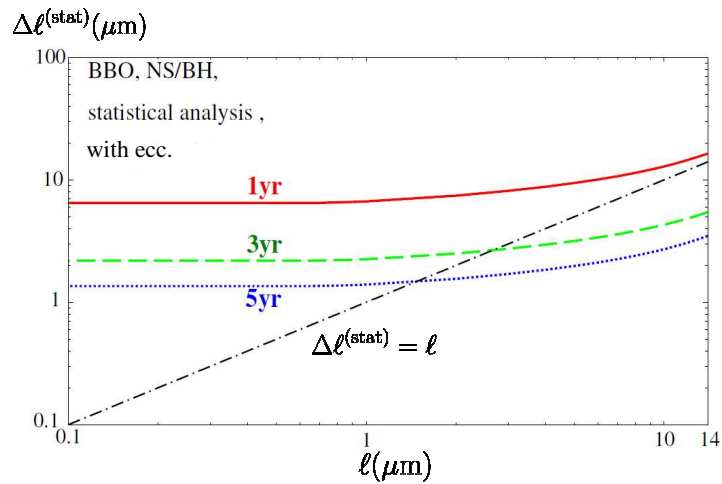} }
 \caption{\label{stat_ecc}
Same as in Fig.~\ref{stat} but here, eccentricity is included in 
binary parameters with the fiducial value $I_e=0$.}
\end{figure}

Next, we consider the effect of eccentricity when we try to measure $\ell$.
We performed the same statistical estimate explained in
Sec.~\ref{sec_stat} but this time, we included $I_e$ into binary
parameters.  
For the actual computation, we set $I_e=0$ as a fiducial value.
The results are shown in Fig.~\ref{stat_ecc}.
Since $\ell$ and $I_e$ have strong degeneracy, 
$\Delta \ell^{(\mrm{stat})}$ becomes
about 5 times larger than the ones shown in Fig.~\ref{stat}.
In this case, 1 yr observation curve does not cross $\Delta \ell^{(\mrm{stat})}=\ell$ line.
This means that when we take into account eccentricity, $\ell$ cannot be
measured with only 1 yr observation period.
However, 3 yr and 5 yr observations will suffice in measuring it with the
detection upper bounds of 2.7$\mu$m (3 yr) and 1.5 $\mu$m (5 yr), respectively. 
These are still 1 order of magnitude stronger than the current table-top result.
These results are also summarized in Table~\ref{table-stat} for GR (third row) and for RS-II (fifth row).

\section{Constraints from the expected numbers of EMRI and BH/NS detection events}
\label{event}

In this section, we derive the constraints on $\ell$ from the expected numbers of events for EMRI with LISA and BH/NS binaries with DECIGO/BBO, extending the discussion  in Sec.~\ref{sec-mc} made by McWilliams~\cite{mc} with LISA.
We do not consider BH/BH binaries with DECIGO/BBO here since these merger rates are expected to be about 1 order of magnitude smaller than the BH/NS ones~\cite{ligo,kalogera2007}. 
We define $\lla N \rra $ as the averaged number of detection events expected in GR and $ \lla N \rra_\mrm{H} $ as the one in RS-II model.
We assume that there is 1 order of magnitude uncertainty in the value of $\lla N \rra $ (see e.g.~\cite{ligo}).
Due to the mass loss effect, $\lla N \rra_\mrm{H} $ is smaller than $\lla N \rra$, and we set the ratio $r_\mrm{H}$ between these two numbers as $\lla N \rra_\mrm{H}=r_\mrm{H}\lla N \rra$ (with $r_\mrm{H} \le 1$).
Since the detection event number follows the Poisson probability distribution~\cite{footnote2}, the variance $\sigma_\mrm{H}^2$ equals the mean $\lla N \rra_\mrm{H}$.
If $\lla N \rra $ events are observed and this value $\lla N \rra $ is within 5-$\sigma$ of the probability distribution, the following inequality,
\beq
\lla N \rra_\mrm{H}+5\sigma_\mrm{H} \ge \lla N \rra,
\eeq
puts the lower bound on $r_\mrm{H}$ as 
\beq
r_\mrm{H} \ge \frac{2\lla N \rra+25 -\sqrt{100\lla N \rra +625}}{2\lla N \rra}. \label{ineq_a}
\eeq

\subsubsection{EMRI Observation with LISA}

LISA can observe a $(5+10^6)\so$ EMRI up to $D_{L,\mrm{(max)}}=4.3$Gpc, assuming that observation starts 1 yr before ISCO and SNR threshold is 10 with 2 effective interferometers. 
Using the EMRI event rate given in Eq.~(\ref{emri}), the averaged number of EMRI detection events $\lla N \rra_\mrm{EMRI}$ becomes
\beqa
\lla N \rra_\mrm{EMRI} &=& \frac{4\pi}{3}D_{L,\mrm{(max)}}^3 T_\mrm{obs} \lla \rch \rra_\mrm{EMRI}  \notag \\
                &\simeq& 3.3\times 10^2 \lmk \frac{D_{L,\mrm{(max)}}}{4.3 \mrm{Gpc}} \rmk^3 \lmk \frac{T_\mrm{obs}}{1\mrm{yr}} \rmk \lmk \frac{M}{10^6\so} \rmk ^{3/8} \lmk \frac{5\so}{m} \rmk ^{1/2}. \label{number_emri}
\eeqa
Then Eq.~(\ref{ineq_a}) becomes $r_\mrm{H}\gsim 0.76$.
This bound is much stronger than the one from the uncertainty in the event rate estimation $\lla N \rra_\mrm{EMRI}$, suggesting that the constraints from Poisson statistics are not appropriate in this case.
We have assumed that the upper bound on $\lla N \rra_\mrm{EMRI}$ within estimation uncertainty will be one order larger than the most probable value.
If 3.3$\times 10^2$ EMRIs are detected as expected with the theoretical upper bound $\lla N \rra_\mrm{EMRI}=3.3\times10^3$, the lower bound on $r_\mrm{H}$ is set as $r_\mrm{H}\gsim 0.1$.
From Eq.~(\ref{rh}), this inequality leads to $\tau \ge 10^9 \mrm{yr}$.
Combining this with Eq.~(\ref{lifetime}), we obtain the upper bound on $\ell$ as
\beq
\ell \lsim 3.9  \lmk \frac{10^9 \mrm{yr}}{\tau} \rmk^{1/2} \lmk \frac{m}{5\so} \rmk^{3/2} \mu \mrm{m}. \label{l-lisa}
\eeq
When only $\lla N \rra=33$ EMRIs are detected with the same theoretical upper bound, the lower bounds on both $r_\mrm{H}$ and $\tau$ are reduced to 10$\%$ of the values mentioned above, leading to $\ell \lsim 12 \mu$m.
These results are summarized in the upper half of Table~\ref{table-event}.

\subsubsection{BH/NS Observation with DECIGO/BBO}

For BH/NS event rate with DECIGO/BBO, we do not have a simple analytic estimate like the one in Eq.~(\ref{emri}).
Therefore we have to rely on the results obtained from population synthesis simulations~\cite{belczynski}. 
Table 2 of Ref.~\cite{belczynski} gives the merger rate of BH/NS binaries as $\rch = 8.1\times 10^{-6} \mrm{MWEG}^{-1} \mrm{yr}^{-1}$ where MWEG stands for the Milky Way Equivalent Galaxy. 
To convert this rate to $\mrm{Mpc}^{-3} \mrm{yr}^{-1}$, we should multiply a factor $ 1.1-1.6 \times 10^{-2} h_{72} \mrm{Mpc}^{-3} $~\cite{kalogera}.
This leads to the BH/NS merger rate of $9.0-13\times 10^{-8}\mrm{Mpc}^{-3} \mrm{yr}^{-1}$.
Following Cutler and Harms~\cite{cutlerharms}, we estimate the averaged number of detection events $\lla N \rra_\mrm{BH/NS}$ for BH/NS binaries as
\beqa
\lla N \rra_\mrm{BH/NS} &=& T_\mrm{obs} \int^{\infty}_{0} 4\pi [a_0 r(z) ]^2\dot{n}(z)\frac{d\tau}{dz} dz \\ \notag
                                     &=& 10^5 \lmk \frac{T_\mrm{obs}}{1\mrm{yr}} \rmk \lmk \frac{\rch}{10^{-7}\mrm{Mpc}^{-1} \mrm{yr}^{-1} } \rmk.
\eeqa
where $a_0 r(z)$ and $d\tau /dz$ are given in Eqs.~(\ref{a0rz}) and~(\ref{dtaudz}), respectively. 
This gives the detection rate of $\lla N \rra_\mrm{BH/NS}=9.0-13\times 10^4$, which yields the 5-$\sigma$ bound from the Poisson statistics as
\beq
r_\mrm{H}\ge 0.98-0.99. \label{ineq_a2}
\eeq
Again, this lower bound is much stronger than the one coming from uncertainty in the theoretical prediction for $\lla N \rra_\mrm{BH/NS}$ and Poisson statistics are inappropriate.
Therefore it is fair to say that if 9.0-13$\times 10^4$ BH/NS binaries are detected, we can only put a constraint $r_\mrm{H} \ge 0.1$ as before.
By using Fig.~\ref{tmerg}, we translate this inequality to the lower bound on $\tau$ as
%
\beq
\tau \gsim 6.9\times 10^7 \mrm{yr}.
\eeq
Using Eq.~(\ref{lifetime}) (or equivalently Eq.~(\ref{l-lisa})), the upper bound on $\ell$ becomes
\beq
\ell \lsim 15 \lmk \frac{6.9\times 10^7 \mrm{yr}}{\tau} \rmk^{1/2} \lmk \frac{M}{5\so} \rmk^{3/2} \mu \mrm{m}, \label{l_belc}
\eeq
where we assumed BH masses to be $M=5\so$.
These results are summarized in the lower half of Table~\ref{table-event}.
We see that the constraints obtained in this section are weaker than the ones with statistical analyses in the previous section.

\begin{table}[t]
\caption{\label{table-event} The constraints on $\ell$ from number of detection events. We also show other important parameters used to derive the final results. }
\begin{center}
\begin{tabular}{c||c|c|c|c}  
 Detectors, binaries, masses &   $\lla N \rra$ & $r_\mrm{H}$ & $\tau$ & $\ell$ \\ 
 &  & &(yr) &  ($\mu$m)  \\ \hline
LISA, EMRI, $(5+10^6)\so$ &  3.3 $\times 10^2$ & 0.1 & $10^9$ & 3.9 \\
&  33 & 0.01 & 10$^8$ & 12 \\ \hline
DECIGO/BBO, BH/NS, $(1.4+5)\so$ &  9.0-13$\times 10^4$ & 0.1 & 6.9$\times 10^7$ & 15  \\ 
\end{tabular}
\end{center}
\end{table}

\section{Conclusions and Discussions}
\label{conclusions}

In this paper, we obtained the possible upper bounds on the size of extra dimension $\ell$ in the RS-II braneworld scenario~\cite{randall2} by detecting GWs from BH/BH and BH/NS binaries with LISA and DECIGO/BBO.
The current table-top experiment puts $\ell \le 14\mu$m~\cite{adelberger}.
It has been conjectured that in the RS-II model, the Hawking radiation is enhanced by a factor depending on $\ell$~\cite{emparan-conj,tanaka-conj}.
This factor becomes 10$^{61}$ for $\ell=10\mu\mrm{m}$ and this mass loss effect may be detected by future GW observations~\cite{mc}.

First, we derived the ``-4PN''-like correction term due to this mass loss effect in the phase of the gravitational waveform.
It is a negative PN order correction which means this mass loss effect is greater when the separation of a binary is larger.
Then we performed Fisher analyses and estimated $\ell_u$, the upper bound on $\ell$.
The constraints from high mass ratio BH/BH observation with LISA are almost independent of the mass of the larger BH $M_0$, while they scale as $m_0^{21/16}$ for the mass of the smaller BH $m_0$.
We gave some analytical explanations to these behaviors.
We found that these constraints using LISA are weaker than the current table-top result.
On the other hand, DECIGO/BBO may put stronger constraint than the table-top experiment by detecting small mass BH/BH and BH/NS binaries.
For example, when GR is the correct theory, 5 yr observation of a (3+3)$\so$ BH/BH binary leads to $\ell \le 3.4\mu$m.
Furthermore, DECIGO/BBO is expected to have very large BH/NS event rate of $O$(10$^5$ yr$^{-1})$.
We found that DECIGO/BBO can measure $\ell$ down to $\ell= 0.33\mu$m for 5 yr observation by performing statistical analysis if we know eccentricities a priori.
This is almost 40 times stronger than the table-top one.
When we include eccentricities into binary parameters, the constraint reduces to $\ell= 1.4\mu$m for 5 yr observation.
When considering actual detection of $\ell$, 
the detection limit becomes $\ell=1.5\mu$m when including eccentricities.
This is still 1 order of magnitude stronger than the table-top one. 
However, table-top experiments are performed model-independently whereas the results obtained here can only apply to RS-II braneworld model.
Therefore we cannot directly compare these results.
Also, since we do not take systematic errors on the waveforms and the limitation of the Fisher analysis into account, calculations in this paper might be underestimating the bounds~\cite{cutler-vallisneri}.
The results are summarized in Table~\ref{table-stat}. 

When performing statistical analyses, we assumed that the distribution of merger time (Fig.~\ref{tmerg}) does not depend on BH mass but this is not true in reality.
The determination error of $\ell$ is mainly determined by the number of smaller mass binaries.
From Fig.~4 of Ref.~\cite{belczynski}, we see that smaller mass binaries have relatively smaller binary separations at their formations.
This is because they obtain large kick velocities when the binary components explode to become compact objects so that only those having small separations can survive.
Since these smaller separation binaries have smaller merger times, it is likely that the peak of the merger time distribution would shift to smaller $t_\mrm{merg}$ when binary BH mass is smaller.
This means that there would be more smaller mass binaries that coalesce within BH evaporation times.
Therefore our estimates shown in Sec.~\ref{sec_stat} are conservative in this sense. 

Next, we derived upper bounds on $\ell$ from the analyses of expected detection event numbers.
We extended the discussions in Ref.~\cite{mc}, including large uncertainties in the event rate estimations.
We found that if the most likely number of EMRIs in GR is detected, LISA can put $\ell \le 3.9\mu$m.
BH/NS binary observations with DECIGO/BBO give 1 order of magnitude weaker constraints.
These results are summarized in Table~\ref{table-event}.


Unfortunately, strong constraint on $\ell$ cannot be obtained from ground-based interferometers.
This is because the optimal frequency range is higher compared to the space-borne ones, meaning that ``-4PN'' correction has less contribution.
To give a concrete example, we estimated $\ell_u$ from the observation of a (3+3)$\so$ BH/BH binary at 1 Gpc with the third generation ground-based interferometer named Einstein Telescope (ET)~\cite{et}.
Its noise spectrum is given in Ref.~\cite{et-noise}.
We found that $\ell_u=5.6\times 10^3\mu \mrm{m}$ with SNR=93, which is much weaker than the one from the table-top experiment.
Constraints are even weaker for the second generation detectors such as advanced LIGO~\cite{ligoweb}, advanced VIRGO~\cite{virgo} and LCGT~\cite{lcgt}.

In this paper, we focused on the correction coming from the mass loss effect.
However, there is another one appearing in the gravitational potential (see Eq.~(\ref{pot})).
In Ref.~\cite{inoue}, Inoue and Tanaka obtained the 2PN correction term to the phase of gravitational waveform due to this potential correction.
We added this 2PN term and performed Fisher analyses but found that our results do not change.  

In this paper, we performed sky-averaged analyses for simplicity.
To take the directions and orientations of the binaries into account, we need to perform Monte Carlo simulations similar to the ones performed in Refs.~\cite{berti,kent,kent2}.   Also, we neglected the effects of BH spins.
The leading contribution of spins appears at 1.5PN with an extra parameter $\beta$, the spin-orbit coupling.
When we add $\beta$ to the binary parameters and perform Fisher analysis for a (1.4+10)$\so$ BH/NS binary at $D_L=3$Gpc with 1 yr DECIGO/BBO observation, we obtain $\ell \le 61\mu$m.
Compared to the one without including $\beta$ (shown in Table~\ref{table-stat} as $\ell \le 51\mu$m), the constraint weakened only slightly.
Since the mass loss correction takes -4PN frequency dependence and spin correction is 1.5PN order, the degeneracy between these two is weak.
Furthermore, when we include the effect of precession, we expect this degeneracy to be solved~\cite{stavridis} so that the inclusion of spin parameter would not affect our results much.
The Hawking radiation also changes the BH spins.
When taking BH spins into account, we may have to consider the spin down effect due to the Hawking radiation.

In reality, the change in the BH mass $M$ is caused not only by the enhanced Hawking radiation but also by the accretion onto the BH.
We need to clarify whether these two effects can be distinguished by Fisher analysis.
The mass loss rate due to the former effect can be expressed from Eq.~(\ref{massloss}) as
\beqa
\dot{M}&=&-\cmdot\left( \frac{\ell}{M} \right)^2 = -5.5\times10^{-9} \left( \frac{10 M_{\odot}}{M} \right)^2 \left( \frac{\ell}{14 \mu \mathrm{m}} \right)^2 M_{\odot} \mathrm{yr}^{-1} \notag \\
          &=&-0.25 \dot{M}_\mrm{edd} \left( \frac{10 M_{\odot}}{M} \right)^3 \left( \frac{\ell}{14 \mu \mathrm{m}} \right)^2,
\eeqa
where $\dot{M}_\mrm{edd}$ is the Eddington accretion rate written as
\beq
\dot{M}_\mrm{edd}=2.2\times 10^{-8} \left( \frac{M}{10 M_{\odot}} \right)   \so \mrm{yr}^{-1}.
\eeq
This shows that if $\ell$ is small, the mass loss effect might be masked by the effect of mass accretion onto the BH if the latter is near Eddington accretion rate.
However, the binaries that suffer from accretion near this rate will be rare. 
Even if the accretion rate is as high as this rate, in general matter accretion goes through accretion disk. 
In this case the conservation of specific angular momentum does not hold and Eq.~(\ref{rhawk}) gets modified, leading to the contribution differing from ``-4PN''.
Therefore we expect that these two effects can be distinguished. 
Also, since $M$ dependence of the mass loss rate is different between these two effects, it may be possible to separate them.
For example, quasi-stationary accretion rate $\dot{M}_\mrm{acc}$ is proportional to the area of the BH yielding $\dot{M}_\mrm{acc}\propto M^2$~\cite{emparan,hoyle,zeldovich,carr} and this has different $M$ dependence compared to Eq.~(\ref{massloss}).
Furthermore, since the properties of BH accretions are intrinsic and not universal, we should be able to reduce these effects statistically.
These issues are left for future works.

\begin{acknowledgments}
\quad We thank Takashi Nakamura, Naoki Seto and Masaki Ando for useful discussions and valuable comments.
K.Y.~is supported by the Japan Society for the Promotion of Science (JSPS) grant No.~$22 \cdot 900$.
N.T.~is supported by the DOE Grant DE-FG03-91ER40674.
T.T.~is supported by JSPS through Grants No.~21244033, the Grant-in-Aid for Scientific Research on Innovative Area Nos.~21111006 and 22111507 from the MEXT.
This work is also supported in part by the Grant-in-Aid for the Global COE Program ``The Next Generation of Physics, Spun from Universality and Emergence'' from the MEXT of Japan. 
\end{acknowledgments}





\begin{thebibliography}{99}
\bibitem{footnote_specific}
When the radiation is emitted isotropically in the rest frame of the BH, there is no additional force.
Therefore the conservation of the specific angular momentum can be obtained by integrating the azimuthal components
of the equations of motion of BHs in the orbital plane (Kepler's second law).
\bibitem{footnote}
This rate is taken from the redshift evolution of NS/NS merger rate obtained in Ref.~\cite{schneider}. Although BH/NS merger rate has not been obtained in this reference, the formation rate of BH/NS follows the same evolution as that of NS/NS (see Fig.~6 of Ref.~\cite{schneider}). Therefore we adopt this evolution for BH/NS merger rate as well.
\bibitem{footnote2}  
McWilliams~\cite{mc} assumed that the EMRI event rate obeys the Poisson statistics, 
but to be more precise, it is the EMRI detection number that obeys these statistics.
However, the discussion made in Ref.~\cite{mc} is valid if we interpret the observation range
 of 1Gpc and the observation time of 1 yr for LISA as implicit assumptions.
\end{thebibliography}
\end{document}